\documentclass[5p,times]{elsarticle}

\usepackage{lineno,hyperref}
\usepackage{url}
\usepackage{hyperref}
\usepackage{doi}
\usepackage{multirow}
\usepackage[linesnumbered,ruled,vlined]{algorithm2e}
\usepackage{amsmath}
\usepackage{mathabx}
\usepackage{booktabs}
\usepackage{svg}
\usepackage{cite}
\usepackage{algorithmic}
\usepackage{graphicx}
\usepackage{textcomp}
\usepackage{utfsym}

\SetCommentSty{mycommfont}
\usepackage{xcolor}
\modulolinenumbers[5]
\usepackage{flushend}
\usepackage{apalike}
\usepackage{pifont}
\usepackage{float}
\usepackage{verbatim} %comments
\usepackage{natbib}
\usepackage{booktabs}

\journal{Journal of \LaTeX\ Templates}

\biboptions{authoryear}

\begin{document}
\begin{sloppypar}
\begin{frontmatter}
	
\begin{titlepage}
	\begin{center}
		\vspace*{1cm}
		
		\textbf{ \large LoadLord: Loading on the Fly to Defend Against Code-Reuse Attacks}
		
		\vspace{1.5cm}
		
		% Author names and affiliations
		Xiaoqi Song$^{a}$ (songxiaoqi@stu.ouc.edu.cn), Wenjie Lv$^a$ (lwj3656@stu.ouc.edu.cn), Haipeng Qu$^a$ (quhaipeng@ouc.edu.cn), Lingyun Ying$^b$ (yinglingyun@qianxin.com)\\
		
		\hspace{10pt}
		
		\begin{flushleft}
			\small  
			$^a$ College of Computer Science and Technology, Ocean University of China, Qingdao 266100, China\\
			$^b$ QI-ANXIN Technology Research Institute, Beijing, China \\
		%	$^c$ Full address of last author, including the country name
			
			\begin{comment}
				Clearly indicate who will handle correspondence at all stages of refereeing and publication, also post-publication. Ensure that phone numbers (with country and area code) are provided in addition to the e-mail address and the complete postal address. Contact details must be kept up to date by the corresponding author.
			\end{comment}
			
			\vspace{1cm}
			\textbf{Corresponding Author:} \\
			Haipeng Qu \\
			College of Computer Science and Technology, Ocean University of China, Qingdao 266100, China \\
			Email: quhaipeng@ouc.edu.cn
			
		\end{flushleft}        
	\end{center}
\end{titlepage}

\title{LoadLord: Loading on the Fly to Defend Against Code-Reuse Attacks}

\author[mymainaddress] {Xiaoqi Song}
\ead{songxiaoqi@stu.ouc.edu.cn}

\author [mymainaddress]{Wenjie Lv}
\ead{lwj3656@stu.ouc.edu.cn}

\author [mymainaddress]{Haipeng Qu$^*$}
\cortext[mycorrespondingauthor]{Corresponding author}
\ead{quhaipeng@ouc.edu.cn}

%\author[mycorrespondingauthor]{Haipeng Qu}

%\author[mysecondaryaddress]{Haipeng Qu}

\author [mysecondaryaddress]{Lingyun Ying}
\ead{yinglingyun@qianxin.com}

% \address{College of Computer Science and Technology, Ocean University of China, Qingdao 266100, China}

\address[mymainaddress]{College of Computer Science and Technology, Ocean University of China, Qingdao 266100, China}
\address[mysecondaryaddress]{QI-ANXIN Technology Research Institute, Beijing, China}

\begin{abstract}
Code-reuse attacks have become a kind of common attack method, in which attackers use the existing code in the program to hijack the control flow. 
Most existing defenses focus on control flow integrity (CFI), code randomization, and software debloating. 
However, most fine-grained schemes of those that ensure such high security suffer from significant performance overhead, and only reduce attack surfaces such as software debloating can not defend against code-reuse attacks completely. 
In this paper, from the perspective of shrinking the available code space at runtime, we propose LoadLord, which dynamically loads, and timely unloads functions during program running to defend against code-reuse attacks. 
LoadLord can reduce the number of gadgets in memory, especially high-risk gadgets. 
Moreover, LoadLord ensures the control flow integrity of the loading process and breaks the necessary conditions to build a gadget chain. 
We implemented LoadLord on Linux operating system and experimented that when limiting only 1/16 of the original function. As a result, LoadLord can defend against code-reuse attacks and has an average runtime overhead of 1.4\% on the SPEC CPU 2006, reducing gadgets by 94.02\%.
\end{abstract}

\begin{keyword}
code dynamically loading \sep code-reuse attack \sep   return-oriented programming \sep buffer overflow \sep control flow integrity
\end{keyword}

\end{frontmatter}

\section{Introduction}
\label{introduction}

Code-reuse attacks are commonly used in exploiting stack buffer overflow vulnerabilities. 
A stack buffer overflow vulnerability allows an attacker to gain control of the instruction pointer, and thus a code-reuse attack can hijack the control flow to a code fragment (usually called a gadget) to gain privilege. 
From the initial return-to-libc\citep{b32} attack, return-oriented programming(ROP)\citep{b30}, and jump-oriented programming (JOP) attacks, there are ways and means of code-reuse attacks, and the protection of computer system security is taken seriously.

Various security mechanisms for code-reuse attacks have been proposed. 
Data execution prevention (DEP)\citep{b33} and address space layout randomization (ASLR)\citep{b31} were deployed in the operating system community. 
The ASLR randomizes offsets in the memory layout, but the segment layout remains the same, this leaves an opportunity for attackers.

For code-reuse attacks, existing research mainly focuses on control flow integrity (CFI)\citep{b34}, code randomization, and software debloating. 
CFI restricts a program to jump within a restricted range through abstract control flow diagrams. Code randomization prevents gadgets from being exploited by modifying the memory code layout. 
Software debloating reduces the number of instruction sequences that may be used by an attacker. 
However, the existing methods still have some drawbacks, such as the need for source code and symbolic debugging information, expensive performance overhead at runtime, and can not describe a complete control flow graph precisely\citep{b12,b17,b44}.

From the perspective of shrinking the available code space in runtime memory, we propose LoadLord to defend against code-reuse attacks using dynamic function loading. 
LoadLord implements dynamic loading of functions, limiting the number of functions loaded into memory during program execution, swapping in and out of loaded functions to ensure that only part of the code is stored in memory, thus reducing the number of gadgets residing in memory, especially gadgets that are vulnerable to exploitation. 
LoadLord can be used as a standalone method and is compatible with existing prevention mechanisms.

LoadLord requires no source information, no debugging information, or symbolic information of the program, and only binary files which are processed by automatic reverse analysis. 
LoadLord loads function dynamically and unloads executed functions promptly to quickly erase information that might be leaked. 
Shrinking the available code space can significantly reduce the number of available gadgets, thereby directly restraining gadget utilization. 
Simultaneously, the loading of a function is restricted as the starting address of the function or the return address of the \verb|call| instruction, which avoids hijacking control flow and ensures the integrity of the control flow in the dynamic loading process.

We implemented the dynamic loading prototype LoadLord, which dynamically loads the required function during program running, limits the number of functions to be loaded, and unloads functions promptly, thus reducing the number of gadgets in the runtime memory. 
The evaluation shows that LoadLord can defend against code-reuse attacks, with a runtime overhead of 1.4\% on the SPEC CPU 2006 and real-world program, and reduces the number of gadgets in runtime memory by 94.02\%.

In this paper, we make the following main contributions:
\begin{itemize}
\item We propose a method of dynamic loading function based on partial control flow integrity, which can dynamically switch functions in and out during program execution to limit the number of functions and reduce the number of gadgets residing in memory.
\item We propose a semantics-based function unload method that breaks the necessary conditions of building a gadget chain by unloading high-risk gadgets.
\item We implemented a dynamic function loading prototype for x86-64 executable files, evaluated the security and performance overhead of our prototype, and found that our prototype generated an average runtime overhead of 1.4\% on the SPEC CPU 2006 benchmark and real-world program, which reduced the number of gadgets in runtime memory by 94.02\%.
\end{itemize}

The remainder of this paper is organized as follows. 
Section \ref{section:back} introduces the background and the threat model. 
Section \ref{section:Overview} introduces the overview of LoadLord. 
Section \ref{section:Implementation} introduces the implementation of LoadLord. 
Section  \ref{section:Evaluation} presents the experiments and provides the results of the evaluation of LoadLord, and Section \ref{section:conclusion} presents the conclusions.

\section{Background and related work}\label{section:back}

\subsection{Related Work}

\begin{figure}[t]
\centerline{\includegraphics[width=0.5\textwidth]{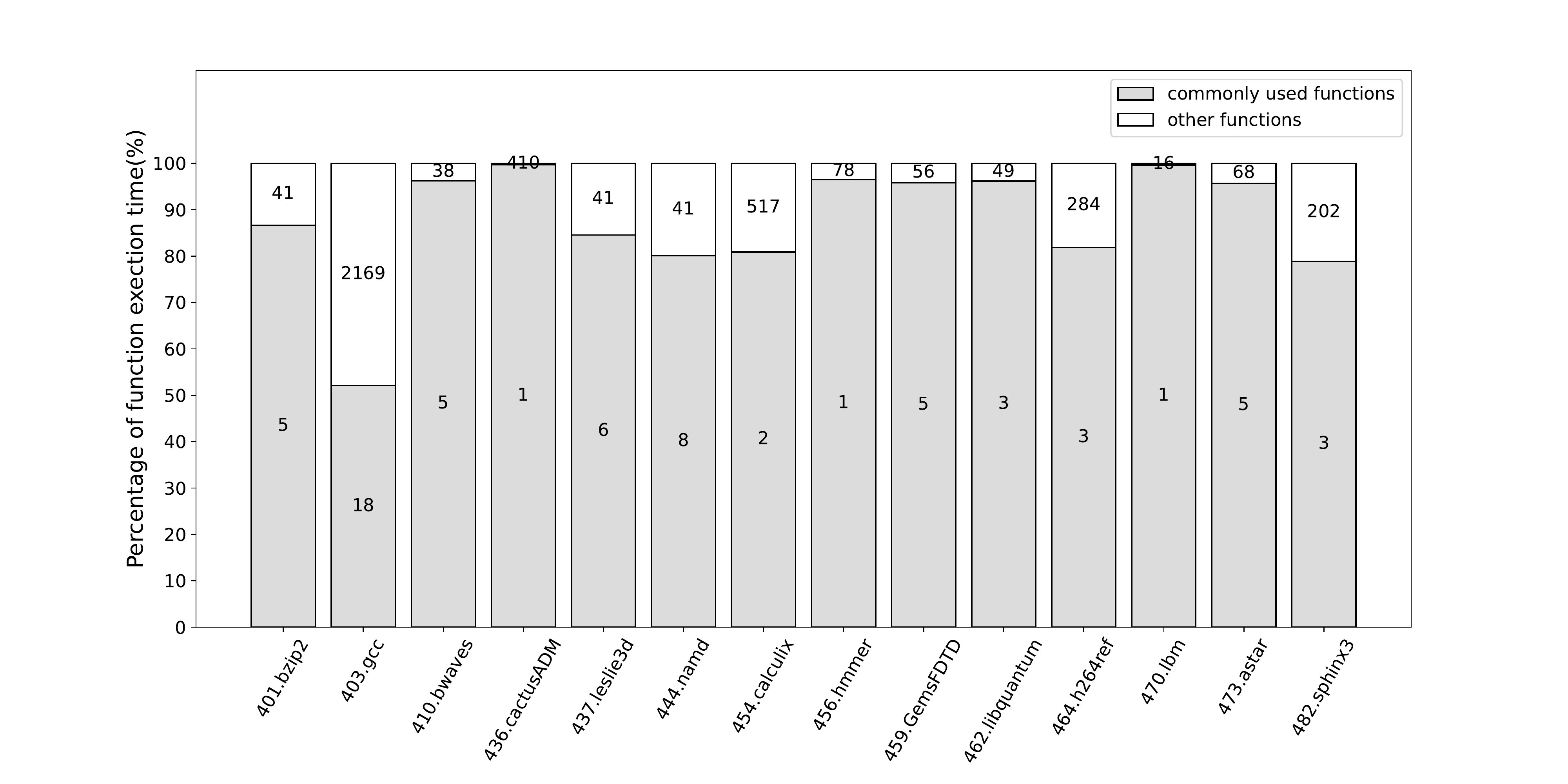}}
\caption{Relationship between program runtime and function runtime}
\label{figmotiv}
\end{figure}
Existing solutions for code-reuse attacks can be divided into three main solutions: CFI, code randomization, and software debloating. 
CFI\citep{b1} prevents control flow hijacking attacks. 
Coarse-grained CFI sacrifices security to decrease overhead\citep{b21}. Fine-grained CFI experiences a high overhead. 
Some studies use the hardware-supported counter Last Branch Record (LBR)\citep{b22, b23,b24,b25} to record branch information, such as indirect jumps and calls, to reduce overhead. 
Some studies reduce the performance overhead of fine-grained CFI by reducing the number of actual validations \citep{b25,b15,b19,b34}. Shadow stack\citep{b26,b27,b35} protects the return address to ensure control flow integrity. Binary CFI mechanisms that do not require source code\citep{b18, b5, b19,b20} are more widely used, but they also have some problems with inaccurate control flow transfer because of the difficulty of accurately characterizing control flow diagrams. 
% Fine-grained\citep{b2,b3,b4} has compatibility problems with complex code\citep{b21}.

Code randomization is another defense mechanism against code-reuse attacks. 
Randomization changes the layout of the address space while the program is loading or running, making the discovered code page layout obsolete and unavailable to gadgets. 
Coarse-grained randomization, such as address space layout randomization, has been deployed in many operating systems. 
Common randomization granularity includes function blocks\citep{b8,b9,b10}, basic blocks\citep{b11, b12,b13}, and instruction and register\citep{b14,b15,b16}. Randomization is implemented with compilers\citep{b11, b12,b28} or binary rewriting tools\citep{b29,b13}. 
These methods make trade-offs on applicability (whether source code is required), disassembly, and jump tracing accuracy and performance\citep{b12}. 
Code randomization requires accurate trace and updating of each pointer and its jump at runtime, incurs costly performance overhead, and the absence of source-code binaries makes analysis difficult.

Software debloating reduces software's attack surface by removing pieces of code that are not required by users. 
Piece-Wise\citep{b47} implements a compile-to-load-time framework to help programs zero out unused shared library parts. 
ROPStravation\citep{b49} makes unused as non-executable through static analysis and fuzzing. 
Nibber\citep{b48} analyzes binary applications and their dependent libraries through static analysis to identify and remove unused library code. 
But software debloating can achieve high gadget count reduction rates, yet fail to limit an attacker’s ability to construct an exploit\citep{b46}.

\subsection{Motivation}
We observed that most functions residing in memory are unused by exploring the relationship between program runtime and functions after using the GNU profiler (\verb|gprof|)\citep{b38} on SPEC CPU 2006\citep{b39}. 
Based on this observation, we propose dynamic function loading to load the function when it is invoked. Because, most of the time, the program relies only on a small number of functions. 
With an appropriate loading limit, the program performance will not be significantly affected, and it can effectively reduce the number of gadgets at runtime.

\verb|Gprof| is a tool that collects timing information of each function. 
We used \verb|gprof| to count the execution time of each function from largest to smallest. As shown in Fig.~\ref{figmotiv}, on average, the most commonly used 6\% of functions make up 90.2\% of the running time of the program.  
A typical example is \verb|cactusADM|, whose most time-consuming function consumes 99\% of the time, and \verb|gcc|'s function execution times are not that different from each other. 

Based on this phenomenon, we propose the idea of loading function dynamically, and design and implement this method.

\subsection{Threat Model}
First, we assume that standard defenses, such as ASLR and position independent executable (PIE) are enabled. 
W$\oplus$ X specifies that no memory address is both writable and executable, and PIE allows the program to relocate the executable to a random location each time it is run. 
Next, we assume that the program contains buffer overflow vulnerabilities, and from the angle of the attacker, we assume that the attacker can use the stack overflow vulnerabilities and leak a pointer of the code to take advantage of the vulnerability. 
% \Ying{Fix me: An attack is a code reuse attack rather than a code injection attack or data, such as LoadLord defense outside the scope of the attack. }
Then, we assume that the attacker only uses code-reuse attack techniques to achieve an attack.
The attacker does not know the program is running under the protection of LoadLord. 
Finally, we focus on only binary executable files without source code, because source code is not always available.

\section{Overview}\label{section:Overview}
\begin{figure}[tbp]
\centerline{\includegraphics[width=0.48\textwidth]{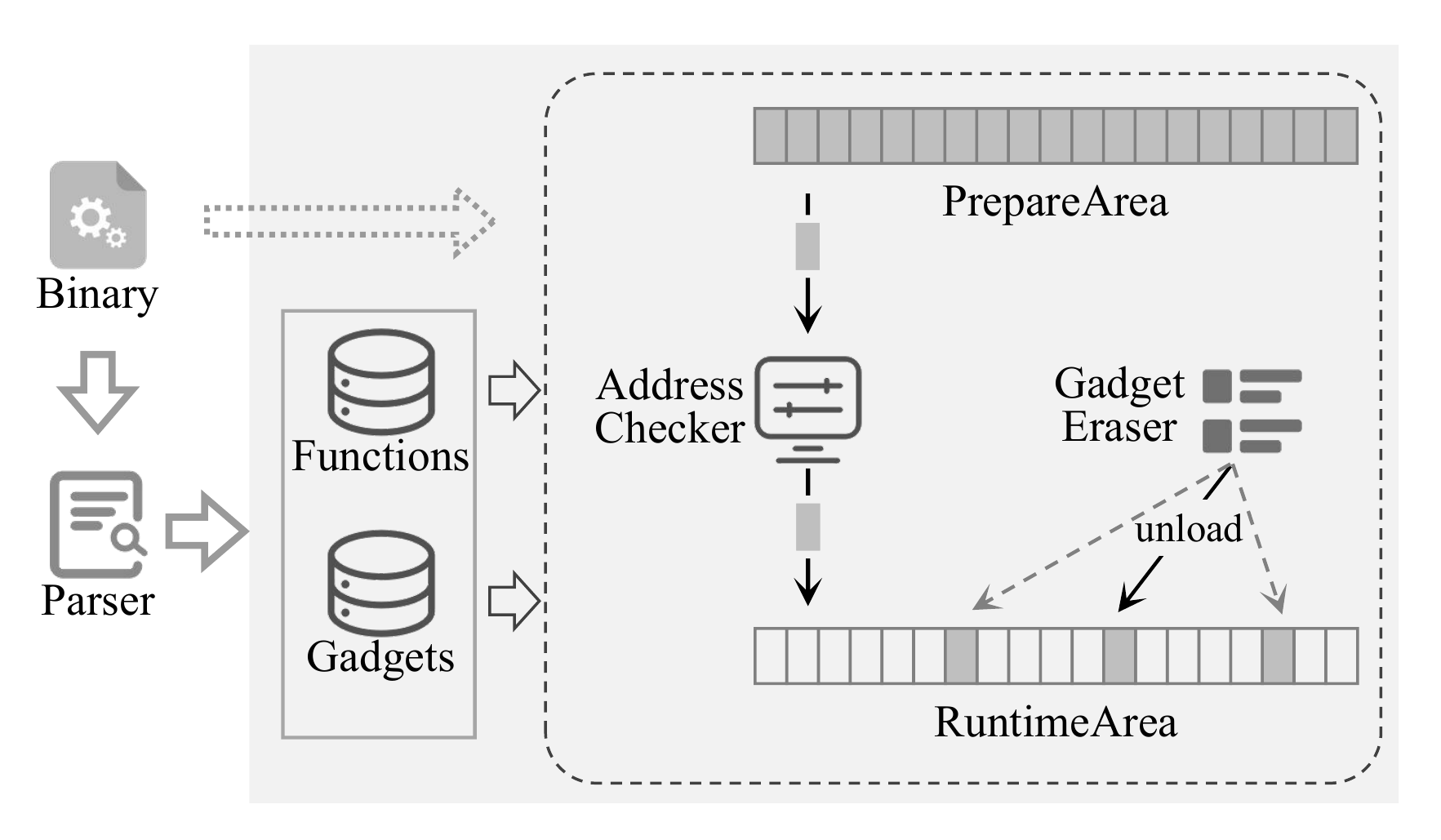}}
\caption{LoadLord architecture}
\label{fig2}
\end{figure}
The LoadLord architecture is shown in Fig.~\ref{fig2}. The \textit{Parser} obtains the segment information, function information, and gadget information from the original binary and determines the information needed for loading. 

The \textit{parser} parses the binary executable file. 
The parser comprises three parts: \textit{Segment Parser}, \textit{Function Parser}, and \textit{Gadget Parser}. \textit{Segment Parser} parses the segment information of the program, analyzes which information needs to be loaded into the memory at runtime, and identifies the location, size, and loading address of each segment. 
The \textit{Function Parser} identifies the starting address and the function size. 
The \textit{Gadget Parser} classifies the gadgets and identifies these gadgets belong to which functions.

LoadLord dynamically loads a program based on its execution and limits the number of loading functions. 
When dynamically loading the functions, the \textit{Address Checker} checks whether the loading address is legal, ensuring control flow integrity during the loading process. 
When unloading functions, the \textit{Gadget Eraser} preferentially unloads functions that are easily exploited, reducing the number of gadgets in memory at runtime, to avoid existing gadgets being exploited.

\section{Implementation}\label{section:Implementation}

\begin{figure*}[tbp]
\centerline{\includegraphics[width=0.8\textwidth]{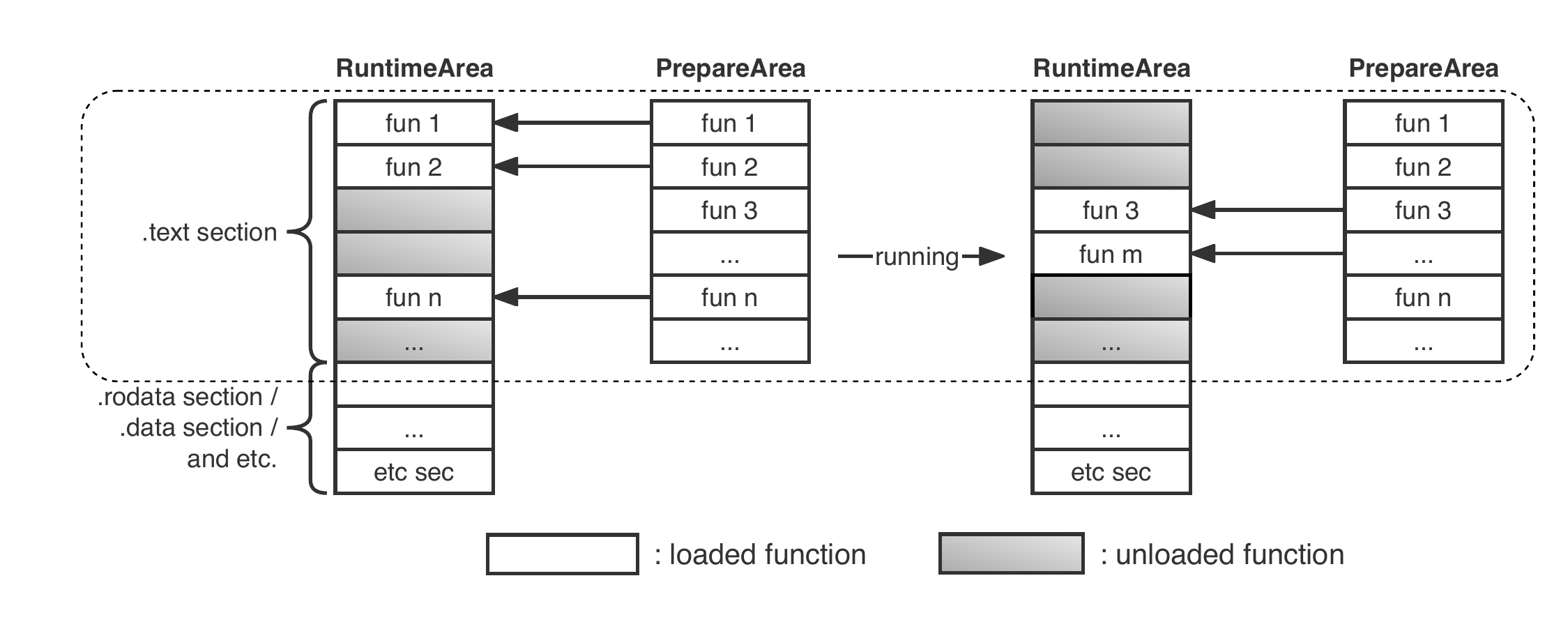}}
\caption{Memory changes during dynamic loading}
\label{fig3}
\end{figure*}
We implemented a dynamic loading prototype, LoadLord, running on Linux x86-64. 
LoadLord supports x86 and x86-x64 binaries without debugging or symbolic information. 
LoadLord follows a three-stage process: pre-processing, dynamic loading functions, and unloading functions, which are discussed in detail as follows.

\subsection{Pre-processing}
The pre-processing stage mainly completes the following tasks:
\begin{table}[htb]
\caption{gadget category}
\begin{center}
\renewcommand\arraystretch{1.4}
\label{tab:gadgettype}
\begin{tabular}{|c|c|c|}
\hline
\textbf{Gadget}   & \textbf{In/Out}   &  \textbf{Define}     \\ \hline
$ArithmeticG$& $InR_1/R_2,OutR$  & $ OutR\leftarrow InR_1 \diamondsuit_b^\mathrm{*} InR_2 $\\ \hline
$LoadMemG$& $AddrR,OutR $ &$ OutR\stackrel{\diamondsuit_b}{\leftarrow}[AddrR+offset]$   \\ \hline
$StoreMemG$& $AddrR,InR$  & $M[AddrR+offset]\stackrel{\diamondsuit_b}{\leftarrow} InR$   \\ \hline
$MoveRegG$& $InR/value,OutR$  & $OutR\leftarrow InR/ value$ \\ \hline
$SYSG$ &-  & $syscall $\\ \hline
$JumpG$ & $AddrR$ & $IP\stackrel{\diamondsuit_b}{\leftarrow}AddrR + offset$ \\ \hline
\multicolumn{3}{l}{$^{\mathrm{*}}$$\diamondsuit_b$ refers to arithmetic and logical operations}
\end{tabular}
\end{center}
\end{table}

\paragraph{Processing segments information} Analyze the segmentation information of binary files and save the offset, size, and permissions of segments that need to be loaded into memory.

\paragraph{Identifying functions} Dynamic loading requires the accurate function identification. 
Based on IDA Pro function recognition, \textit{Function Parser} divides functions more accurately according to the \verb|call| instructions.

% \textit{Function Parser} regards the destination address of each call instruction as a function entry address and analyzes all call instructions to distinguish functions with maximum accuracy. 
When identifying the functions, \textit{Function Parser} first uses IDA Pro to identify the functions, and then checks the identification results for each \verb|call| instruction of IDA to determine whether the destination address is a start address for a function; if not, \textit{Function Parser} divides the destination function into two pieces, the destination address as the starting address of the new partition function.

\begin{figure}[htbp]
\centerline{\includegraphics[height=3.1cm]{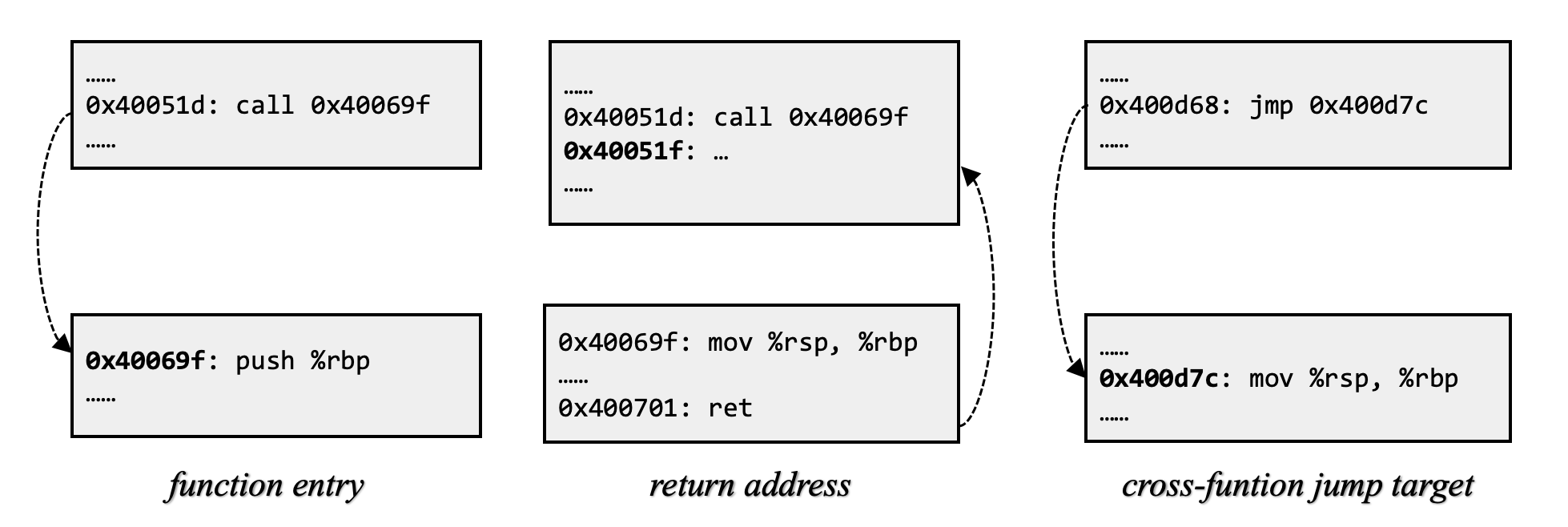}}
\caption{Three legal addresses}
\label{figaddr}
\end{figure}

\paragraph{Identifying legal addresses} Three types of legal addresses are defined to ensure the integrity of partial control flow in the loading process, as shown in Fig.~\ref{figaddr}: the starting address of the function, return address of the \verb|call| instruction, and target address of the \verb|jmp| instruction which will jump to another function. 
These three addresses correspond to the three types of functions that need to be loaded during the operation:

\begin{itemize}
\item Load a new function.
\item The function has unloaded when the callee function returns.
\item Jump to another function.
\end{itemize}

Considering that in a normal control flow, a function can only be entered or returned through these three types of addresses, we use these three types as legal addresses to ensure that dynamically loaded functions are valid.

\paragraph{Identifying and classifying gadgets} In order to avoid the gadgets that reside in memory constructed as highly available chains, we analyze the conditions for chain construction and classify the gadgets.
Necessary gadgets needed to construct a chain are prioritized during function unloading.
First, we identify gadgets in the program and classify existing gadgets functionally. 
Table.~\ref{tab:gadgettype} shows the functional categories of the gadgets, which are divided into \textit{ArithmeticG, LoadMemG, StoreMemG, MoveRegG, SYSG, JumpG}. 
Distinguish the functions according to these gadget types.

\subsection{Dynamic Function Loading}
LoadLord parses the pre-processing information, saves the starting address and size of each function, and return address of each \verb|call|. 
Then, LoadLord forks a child process and transfers control flow to the entry address.
LoadLord dynamically loads functions while the program runs. 
LoadLord first allocates memory to the program based on the pre-processing information. 
The memory allocation for the program can be divided into two types. 
One type contains the backup of the code segment of the program, which is marked as read-only and is called \textit{PrepareArea}. The second is organized according to the original binary layout, with the code segments cleared, called \textit{RuntimeArea}. LoadLord copies functions from \textit{PrepareArea} to \textit{RuntimeArea} based on program running requirements. 
When functions need to be loaded, LoadLord dynamically marks the area as writable and immediately becomes readable and executable after the function is loaded. 
As shown in Fig.~\ref{fig3}, the function loaded by the program at a certain moment include \verb|fun_1|, \verb|fun_2|, and \verb|fun_n|, that is, these functions run in memory at the current moment. 
As the program runs, at a later point, LoadLord unloads \verb|fun_1|, \verb|fun_2|, and \verb|fun_n|, loads \verb|fun_3| and \verb|fun_m| into the \textit{RuntimeArea}. 
This is how LoadLord keeps the program running by switching the function in and out.

When LoadLord receives a \verb|SIGTRAP| signal, that means the child process steps in a function which not been loaded in memory. 
An interrupt signal can occur either when it steps into an unloaded function or when it returns to an unloaded function. 
In the first case, LoadLord obtains the function information from the function map and loads it; in the second case, LoadLord uses static analysis to record the return address of all \verb|call| instructions and determines which function needed to be reloaded by the current return address.
\subsection{Semantics-based Function Unloading}
\begin{figure}[tbp]
\centerline{\includegraphics[height=3.1cm]{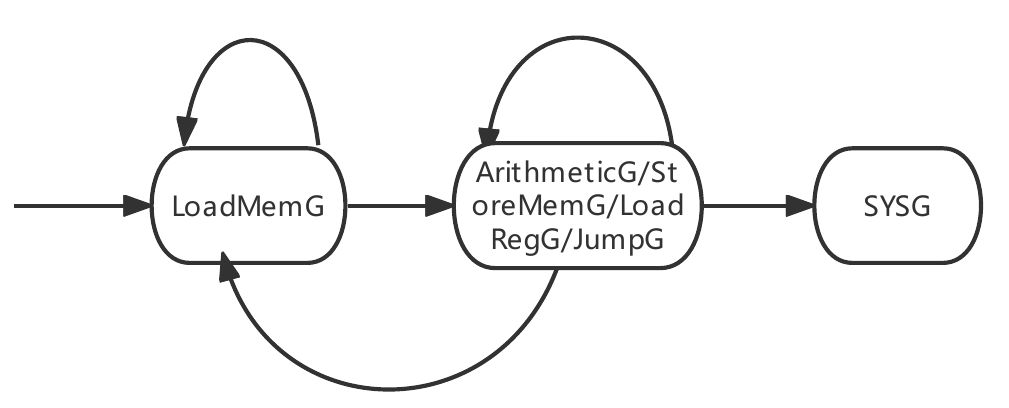}}
\caption{Basic rules for constructing a gadget chain}
\label{figchain}
\end{figure}
Function unloading is triggered in two ways: the number of loaded functions reaches the upper limit or the loaded functions can build a gadget chain. 

The number of loaded functions is checked each time a function is loaded. 
When the number of loaded functions reaches a certain upper limit, LoadLord unloads functions according to the first-in-first-out principle. 

Consider the conditions of destroying the construction of an existing gadget chain when unloading the function. 
First, we analyze the existing gadgets in memory and abstract the basic rules for constructing an effective gadget chain, 
as shown in Fig.~\ref{figchain}, the \textit{LoadMemG} type is necessary to construct the gadget chain. 
We generalize \textit{LoadMemG} into two forms: \verb|pop reg| and  \verb^mov|sub|add reg, [Reg + offset]^. 
Unload the functions containing these types of gadgets in time, so the necessary gadgets are missing during the process running, 
that the attacker cannot master enough knowledge to implement a complete attack.

\begin{figure*}[ht]
\centerline{\includegraphics[width=\textwidth]{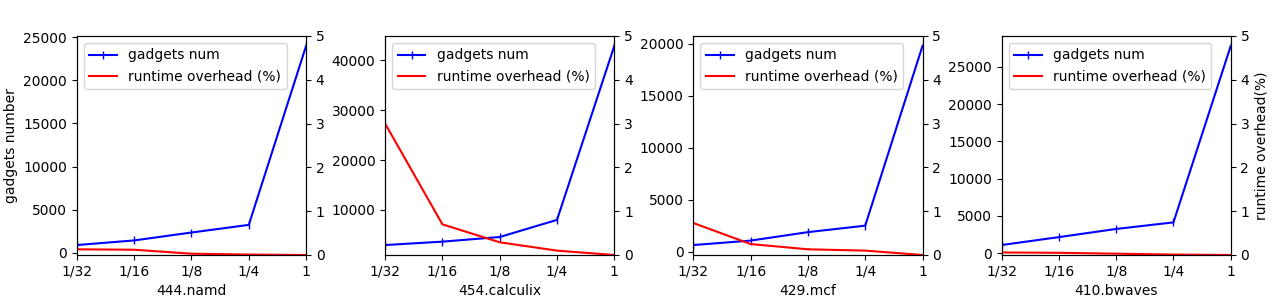}}
\caption{Run overhead and gadget number changes under different loading limits}
\label{fig4}
\end{figure*}

\section{Evaluation}\label{section:Evaluation}
\subsection{Experimental Setup}
In this section, we present the experimental evaluation results of the LoadLord prototype in four parts: correctness, gadget elimination, effectiveness, and performance. 
First, LoadLord does not change the normal operation of the program, thus preserving the original semantics of the binary program. 
Second, LoadLord successfully defended against ROP attacks. 
In addition, LoadLord reduced the number of gadgets running in the program by 94.02\%. Finally, we estimated that the runtime overhead for LoadLord is 1.4\%. Experiments were conducted on a Ubuntu 20.04 LTS, with a 3.10 Hz 8 cores CPU and 16 GB RAM. 
We evaluated the correctness using automation tools ROPgadget\citep{b37} and Ropper\citep{b36} and the performance overhead using the SPEC CPU 2006 Benchmark because it contains many applications with different scenarios and other related work uses it.
\subsection{Correctness}
To ensure that LoadLord does not break the program, we run some applications, such as the executable files in the SPEC CPU 2006 Benchmark and real-world program, and verified that they are opened and executed correctly. 
Dynamic loading does not need to modify the source code or compiler, and maintains the original semantics of the binary program.
\subsection{Gadget Elimination}
The loading limit must be selected carefully because it directly affects both the security and performance. 
If the loading limit is too large, LoadLord may remain sufficient to complete an attack. 
Conversely, if the loading limit is too low, it will cause a large performance overhead. 
We considered the impact of the loaded function number on the runtime overhead. 
In the dynamic loading stage, the number of loading functions is negatively correlated with the time cost, the lower the limit loading functions, the higher the time cost. 
To reduce the impact of the load runtime variability, we tested each application 100 times. 
With the different limits of the load function number (1/4, 1/8, 1/16, 1/32 of the total), we obtained Fig.~\ref{fig4}. 
Considering that computing the minimum number of functions required for large/small programs to run is complex, and unloading functions keep loaded functions safe, we set the upper limit to be proportional instead of a fixed number. Combining the two factors of security and performance, we will load the limit to 1/16.

The most significant effect of dynamic loading is to reduce the number of gadgets in the memory, which directly affects the generation of payload. 
We used the static gadget detection tool, Ropper, to measure the number of gadgets in the raw binary and the number of gadgets in memory during dynamic loading while the program is running. 
Table.~\ref{tab1} shows that when the loading function is limited to 1/16, dynamic loading during program execution reduces the number of gadgets in the memory by 94.02\% on average, and significantly reduces the number of \textit{LoadMemG}, which is necessary to build a gadget chain.

\begin{table}[ht]
\caption{The number of gadgets reduced at runtime}
\begin{center}
\renewcommand\arraystretch{1.2}
\begin{tabular}{|c|c|c|c|c|}
\hline
~ &\textbf{\# of } &\textbf{\# of   }&~&\textbf{\# of }\\
~ &\textbf{Total} &\textbf{Survive  }&\textbf{Reduce}&\textbf{ Load-}\\
\textbf{Benchmark} &\textbf{ Gadget} &\textbf{ Gadget }&\textbf{(\%)}&\textbf{MemG}\\
%\textbf{Table}&\multicolumn{3}{|c|}{\textbf{Table Column Head}} \\
%\textbf{Head} & \textbf{\textit{Table column subhead}}& \textbf{\textit{Subhead}}& \textbf{\textit{Subhead}} \\
\hline
bzip2 &20,058 &925 &95.39& 52\\ \hline
bwaves &27,791 &2,164 &92.21&38 \\ \hline
mcf &19,800 &1,085 &94.52& 49\\ \hline
namd &23,974 &1,443 &93.98&64\\ \hline
calculix &42,839 &3,658 &91.46&75\\ \hline
hmmer &28,125 &1,112 &96.05&94\\ \hline
GemsFDTD &32,118 &1,813 &94.36&97\\ \hline
tonto &61,125 &3,052 &95.01&217\\ \hline
astar &22,213 &492 &97.79&35\\ \hline
cactusADM &41,868 &3,638 &91.42&146\\ \hline
gobmk &32,468 &1,742 &94.74&63\\ \hline
povray &45,144 &3,137 &93.06&114\\ \hline
lbm  &17,056 &1,174 &93.12&33\\ \hline
adpcm &16,644 &1,005 &93.97&43 \\ \hline
CRC32 &16,449 &963 &94.15&50 \\ \hline
rsynth &21,552 &1,494 &93.12&83\\ \hline
\textbf{Average} &- &- & \textbf{94.02} & - \\
\hline
\end{tabular}
\label{tab1}
\end{center}
\end{table}

\begin{table*}[htbp]
\caption{Effectiveness for Different Attack Methods}
% \begin{center}
% \small
{
\renewcommand\arraystretch{1.3}
\setlength{\tabcolsep}{3.9mm}
\begin{minipage}{\textwidth}
\vspace{3mm}
\begin{minipage}{0.5\textwidth}
\begin{flushleft}
\begin{tabular}{|c|c|c|c|c|c|}
\hline
% ~ &\textbf{Overflow} &\textbf{Attack}&\textbf{Target}&~&\textbf{Defen-}\\
% \textbf{No.} &\textbf{Technique} &\textbf{Code}&\textbf{Code Pointer}&\textbf{LOC}&\textbf{-se?}\\ \hline
\textbf{No.} &\textbf{O.T.} &\textbf{ATK.}&\textbf{Target}&\textbf{LOC.}&\textbf{DF.?}\\
\hline
1 & D &cf& RET & stack &\ding{52} \\ \hline
2 & D &cf& BP & stack &\ding{52} \\ \hline
3 & D &cf& FPSP & stack &\ding{52}\\ \hline
4 & D &cf& FPSV & stack &\ding{52}\\ \hline
5 & D &cf& LJSV & stack &\ding{52}\\ \hline
6 & D &cf& sFPS & stack &\ding{52}\\ \hline
7 & D &cf& RET & heap &\ding{52}\\ \hline
8 & D &cf& FPSV & heap &\ding{52}\\ \hline
9 & D &cf& FPH & heap &\ding{52}\\ \hline
10 & D &cf& LJH & heap &\ding{52}\\ \hline
11 & D &cf& sFPH & heap &\ding{52}\\ \hline
12 & D &cf& sFPD & heap &\ding{52}\\ \hline
13 & D &cf& FPB & bss &\ding{52}\\ \hline
14 & D &cf& LJB & bss &\ding{52}\\ \hline
15 & D &cf& sFPB & bss &\ding{52}\\ \hline
16 & D &cf& FPD & data &\ding{52}\\ \hline
17 & D &cf& LJD & data &\ding{52}\\ \hline
18 & D &cf& sFPD & data &\ding{52}\\ \hline
% 19 & D &r2l& RET & stack &\ding{52}\\ \hline
% 20 & D &r2l& BP & stack &\ding{52}\\ \hline
% 21 & D &r2l& FPSV & stack &\ding{56}\\ \hline
% 22 & D &r2l& FPSP & stack &\ding{56}\\ \hline
% 23 & D &r2l& LJSV & stack &\ding{52}\\ \hline
% 24 & D &r2l& sFPS & stack &\ding{56}\\ \hline
% 25 & D &r2l& FPH & heap &\ding{52}\\ \hline
% 26 & D &r2l& LJH & heap &\ding{56}\\ \hline
% 27 & D &r2l& sFPH & heap &\ding{56}\\ \hline
% 28 & D &r2l& FPB & bss &\ding{52}\\ \hline
% 29 & D &r2l& LJSV & bss &\ding{52}\\ \hline
% 30 & D &r2l& LJB & bss &\ding{52}\\ \hline
% 31 & D &r2l& FPD & data &\ding{52}\\ \hline
% 32 & D &r2l& LJD & data &\ding{52}\\ \hline
% 33 & D &r2l& sFPD & data &\ding{56}\\ \hline
19 & D &rop& LJD & data &\ding{52}\\ \hline
20 & D &rop& sFPS & data &\ding{52}\\ \hline
21 & D &rop& sFPB & data &\ding{52}\\
\hline
22 & ND &cf& RET & stack &\ding{52} \\ \hline
23 & ND &cf& BP & stack &\ding{52} \\ \hline
24 & ND &cf& FPSP & stack &\ding{52}\\ \hline
25 & ND &cf& FPSV & stack &\ding{52}\\ \hline
26 & ND &cf& FPH & stack &\ding{52}\\ \hline
27 & ND &cf& FPB & stack &\ding{52}\\ \hline
28 & ND &cf& FPD & stack &\ding{52}\\ \hline
\end{tabular}
\end{flushleft}
\end{minipage}
\begin{minipage}{0.5\textwidth}
\begin{flushleft}
\begin{tabular}{|c|c|c|c|c|c|}
\hline
\textbf{No.} &\textbf{O.T.} &\textbf{ATK.}&\textbf{Target}&\textbf{LOC.}&\textbf{DF.?}\\
% ~ &\textbf{Overflow} &\textbf{Attack}&\textbf{Target}&~&\textbf{Defen-}\\
% \textbf{No.} &\textbf{Technique} &\textbf{Code}&\textbf{Code Pointer}&\textbf{LOC}&\textbf{-se?}\\ \hline
\hline

29 & ND &cf& LJSV & stack &\ding{52}\\ \hline
30 & ND &cf& LJSP & stack &\ding{52}\\ \hline
31 & ND &cf& LJH & stack &\ding{52}\\ \hline
32 & ND &cf& LJB & stack &\ding{52} \\ \hline
33 & ND &cf& LJD & stack &\ding{52} \\ \hline
34 & ND &cf& RET & bss &\ding{52} \\ \hline
35 & ND &cf& BP & bss &\ding{52} \\ \hline
36 & ND &cf& FPSP & bss &\ding{52}\\ \hline
37 & ND &cf& FPSV & bss &\ding{52}\\ \hline
38 & ND &cf& FPH & bss &\ding{52}\\ \hline
39 & ND &cf& FPB & bss &\ding{52}\\ \hline
40 & ND &cf& FPD & bss &\ding{52}\\ \hline
41 & ND &cf& LJSV & bss &\ding{52}\\ \hline
42 & ND &cf& LJSP & bss &\ding{52}\\ \hline
43 & ND &cf& LJH & bss &\ding{52}\\ \hline
44 & ND &cf& LJB & bss &\ding{52} \\ \hline
45 & ND &cf& LJD & bss &\ding{52} \\ \hline
46 & ND &cf& RET & data &\ding{52} \\ \hline
47 & ND &cf& BP & data &\ding{52} \\ \hline
48 & ND &cf& FPSP & data &\ding{52}\\ \hline
49 & ND &cf& FPSV & data &\ding{52}\\ \hline
50 & ND &cf& FPH & data &\ding{52}\\ \hline
51 & ND &cf& FPB & data &\ding{52}\\ \hline
52 & ND &cf& FPD & data &\ding{52}\\ \hline
53 & ND &cf& LJSV & data &\ding{52}\\ \hline
54 & ND &cf& LJSP & data &\ding{52}\\ \hline
55 & ND &cf& LJH & data &\ding{52}\\ \hline
56 & ND &cf& LJB & data &\ding{52} \\ \hline
% 72 & ND &cf& LJD & data &\ding{52} \\ 
% \hline
\end{tabular}
\end{flushleft}
\end{minipage}
\vspace{2mm}
\end{minipage}
}
% \end{center}
{
\footnotesize
% % \linespread{0.2mm}
% % \setlength{\baselineskip}{2pt}
% % \begin{spacing}{0.7}
NOTE: ATK = attack method, O.T. = overflow technology, LOC = attack memory location, DF? = if the defense success, D = direct, ND = indirect, cf = creatfile, r2l = return to libc, RET = ret, BP = baseptr, FPSP = funcptrstackparam, FPSV = funcptrstackvar, LJSV = longjmpstackvar, sFPS = structfuncptrstack, FPH = funcptrheap, LJH = longjmpheap, sFPH = structfuncptrheap, sFPD = structfuncptrdata, FPB = funcptrbss, LJB = longjmpbss, sFPB = structfuncptrbss, FPD = funcptrdata, LJD = longjmpdata, sFPD = structfuncptrdata, 
% % \end{spacing}
}
\label{figripe}
\end{table*}
\subsection{Effectiveness}
To estimate its effectiveness against code-reuse attacks, we tested our approach against the open-source RIPE benchmark\citep{b45}. The RIPE benchmark consists of 3,840 attacks, of which 876 can succeed on our platform. 
The results are shown in Table.~\ref{figripe}, ``Overflowing Technique'' means \textit{direct} or \textit{indirect} overflowing techniques, ``Attack Code'' means spawns a shell on the vulnerable machine or creates a file in a specific directory, ``Target Code Pointer'' is the code pointer to redirect towards the attack code, ``Location'' is the memory location of the buffer to be overflowed. 
LoadLord could detect 796 of these attacks, and the remaining 80 attacks were return-to-libc attacks, which call a library function directly. 
We admit that such attacks can escape because the call destination is a function entry, which is a normal control flow in our method.

LoadLord limits the number of functions that have been loaded into the memory during the dynamic loading of binary programs. 
As a result, only a limited number of gadgets exist in memory. 
As the program runs, the functions loaded in the memory are constantly updated, and the detected gadgets become unavailable at the next moment.

A set of gadgets that generates a payload is critical to an attack, and the effectiveness of a gadget chain generated from a single gadget can be determined. 

\begin{figure*}[t]
\centerline{\includegraphics[width=0.95\textwidth]{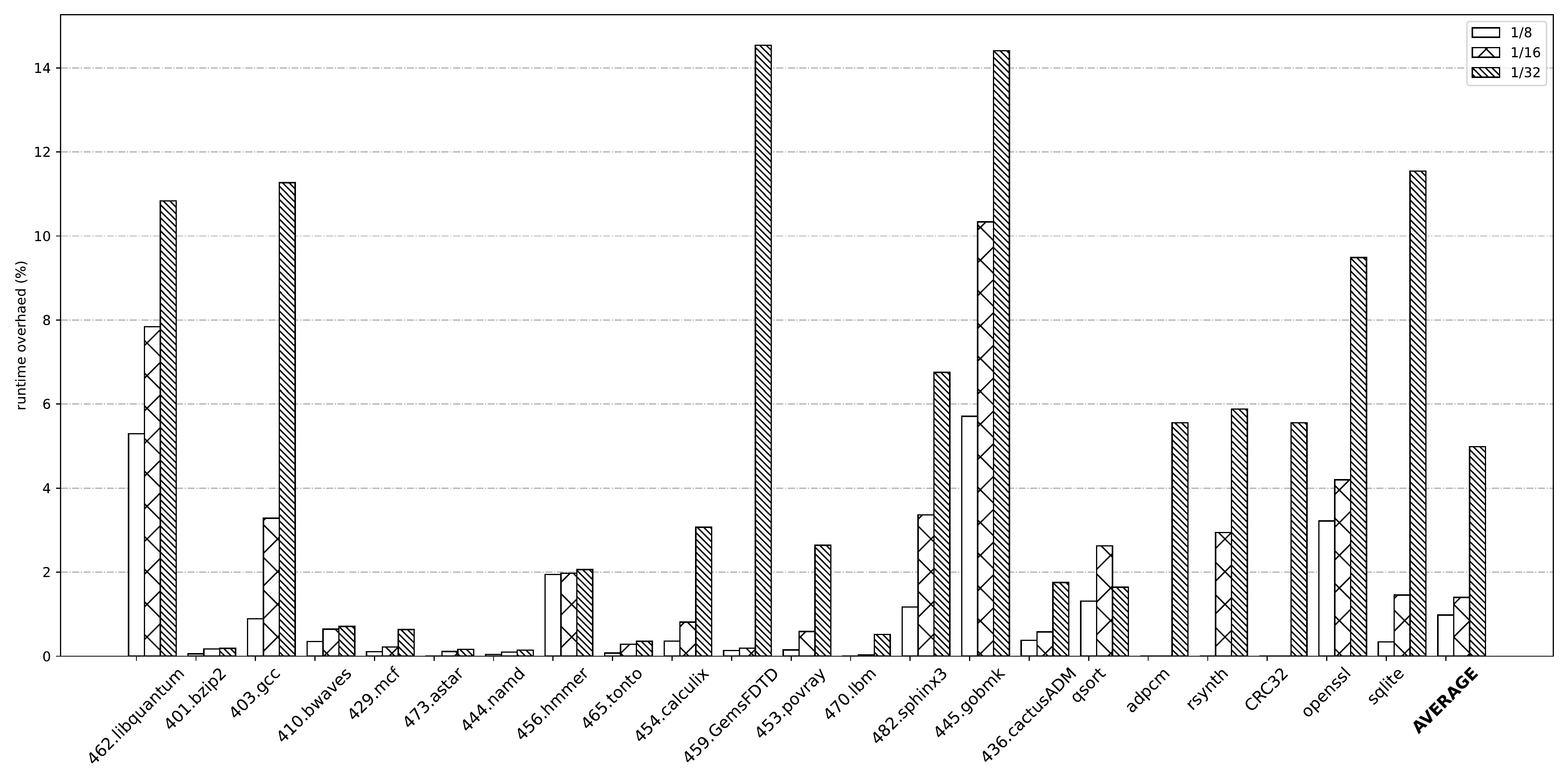}}
\caption{Runtime overhead for LoadLord}
\label{fig5}
\end{figure*}
We used the static gadget detection and payload generation tool, ROPgadget, to evaluate the validity of the gadgets. 
When the dynamically loaded program cannot construct the payload, or when the generated payload cannot complete the attack in the dynamically loaded program, the payload is considered invalid, and LoadLord successfully defends against the attack. 
ROPgadget can construct a payload for binary programs in the SPEC CPU2006 benchmark, indicating that these programs have vulnerabilities that can be exploited. 
During the program, the ROPgadget cannot construct a payload for random memory dump operations, especially for input operations. 
LoadLord is an effective defense against attacks.

\subsection{Performance}
We evaluated the efficiency of LoadLord by measuring program runtime overhead using the SPEC CPU 2006 Benchmark and real-world programs for performance evaluation.

In the dynamic function loading, whenever a new function is loaded, LoadLord interacts with the underlying OS to interrupt it. 
Accordingly, excessive loading degrades the performance of LoadLord. 
Therefore, the load number must be set to a proper value to prevent the overhead from increasing.

We ran each binary file in the benchmark 100 times to eliminate runtime uncertainty. 
Fig.~\ref{fig5} shows the runtime overhead for LoadLord. 
We ran three sets of experiments, with the number of functions loaded as variables. 
The loading functions are limited to 1/8, 1/16, and 1/32 of the original number of functions, respectively. 
At the 1/16 limit, the average overhead is 1.4\%. 

LoadLord overhead is mainly produced in dynamic loading when the interrupt processing program runs most of the time to perform only a finite number of functions. 
After loading the frequently used function, the program uses a function loaded in memory more than loads a new function commonly, which makes the load cost under most load limits too large.

\section{Conclusion and future work}\label{section:conclusion}
\subsection{Conclusion}
In this paper, we present a new approach called LoadLord to effectively defend against code-reuse attacks, which can dynamically load functions to reduce the number of gadgets in memory and unload functions in time to break the necessary conditions to build a gadget chain. LoadLord guarantees partial control flow integrity by restricting the loading from the start address, although this is not effective for return-to-libc attacks. 
We implement a dynamically loading prototype that requires no modifications to the compiler or kernel, nor access to source code, just operates on binary executables. 
LoadLord eliminates 94.02\% of the gadgets on average and incurs an average overhead of 1.4\%.
\subsection{Future Work}
In the next step, this paper will explore the defense effectiveness of more variant attacks. This paper implements the LoadLord prototype through the dynamic loading function scheme and the dynamic link library permission control scheme, and analyzes its defense effectiveness against ROP, JOP and return to libc, but there are still some variant attack techniques that are not in the scope of this paper. , such as the JIT-ROP attack technology, can realize this type of attack when the script interaction that supports just-in-time compilation is supported in the browser. In the future, we will further improve the applicability of this solution and strive to provide protection in more application scenarios.

%\section*{Acknowledgements}

% This work was supported by the National Natural Science
% Foundation of China [grant number 61827810].

\bibliography{ref1}
\end{sloppypar}
\end{document}